\documentclass[prl,twocolumn,showpacs,preprintnumbers,amsmath,amssymb]{revtex4} 
\usepackage{epsfig}
\usepackage{dcolumn}
 \usepackage{bm}
 
\begin{document}
\preprint{
\hfill  TH01.X
} 

\title{
Zone Edge Softening and Relaxation in the Double Exchange Model.
} 

\author{S. E. Barnes and S. Maekawa }

\affiliation{Institute for Materials Research, Tohoku University, 
Sendai 980-8577, Japan}

\date{\today} 
\begin{abstract}
{The $J\to \infty$ double exchange model is formulated in terms of 
three auxiliary particles.  A slow true bosonic magnon propagates by 
admixture with a fast fermionic pseudo-magnon.  This process involves 
the absorption of a conduction electron which, for this half-metal, 
carries only charge degrees of freedom.  The magnon dispersion becomes
much weaker and the relaxation rate increases rapidly upon approaching the 
zone boundary.  That the magnons relax for all wave vector values 
implies the existence of a low energy spin continuum.  }
\end{abstract}

\maketitle

The double exchange model has a long history.  It was introduced by 
Zener\cite{1} fifty years ago in order to explain the phase diagram of 
La\(_{1-x}\)Ca\(_{x}\)MnO\(_{3}\).  With the explosion of interest in 
the colossal magnetoresistance effect in a number of manganite systems 
this has again become a model of prime importance. The Hamiltonian is,
\begin{equation}
{\cal H} = 
- J \sum_{i} \vec S_i\cdot \vec s_i - \sum_{\sigma nn^{\prime}}
t_{nn^{\prime}}(c^\dagger_{n\sigma} c^{}_{n^{\prime}\sigma} + H.c.).
 \label{un}
 \end{equation}
Here $J$ reflects the Hund's rule coupling energy and is taken to be 
positive and large, i.e., $J> t$.  In a seminal paper, de 
Gennes\cite{2,3} showed for near neighbor $t_{nn^{\prime}}$ couplings 
that in the classical, i.e.,large spin \(S\), limit this model has 
magnon like excitations \(\propto (x t/S) (1-\cos q)\), i.e., the 
dispersion relationship is that of a Heisenberg model.  For such large 
values of $J$ the model is a half-metal.  The Fermi level is in the 
majority band and Stoner excitations have large energies $\sim J$ and 
are well above the much lower energy magnons.

The experimental situation has been discussed recently\cite{Dai} by 
Dai et al.  Materials with high Curie temperatures {\it do\/} have 
Heisenberg like dispersion relationships {\it but\/} for similar 
compounds with smaller such temperatures there is a marked softening 
of the magnons upon approaching the zone boundary.  The relaxation 
rate is also anomalous.  The magnon width is quite small for small 
$\vec q$ increasing very strongly for $q$ larger than some critical 
value.  Since this behavior is not explained by double exchange model 
using the standard theory based upon the large $S$ approach nor by the 
theory of magnon-magnon interactions, these authors suggest that the 
coupling to phonon modes is important.  The theoretical understanding 
has been summarized\cite{KK} by Khaliullin and Kilian.  They also 
emphasize that the experimental situation cannot be explained by the 
usual theory and develop a large $S$ approach which includes coupling 
to Jahn-Teller phonons and which {\it can\/} explain experiment.

In this Letter will be presented a novel analytic theory of the basic 
$\cal H$, Eqn.~(\ref{un}), which shows that the two experimental 
effects attributed to phonons do in fact occur within the simple 
double exchange model.  This new approach is based upon the auxiliary 
particle method.  There are three such particles, a $b$-boson which 
reflects the true, slow, magnons, the $c$-fermion which represents the 
charge degrees of freedom of the conduction electrons and a $t$-fermion 
which corresponds to a spin deviation at a site with a charge carrier.

In fact, the earlier exact diagonalization studies of Zang et al., 
\cite{4} and Kaplan and Mahanti \cite{5} for the one dimensional model 
indicate the existence of zone boundary softening.  In particular for 
\(S=1/2\) and small concentrations the magnon dispersion deviates 
strongly from \((1-\cos q)\).  Very recently Kaplan et al., \cite{7} 
have performed new exact diagonalization studies again for one 
dimension and in the limit $J \to \infty$.  Contrary to popular belief 
and mean field theory, they find spin excitations with energies much 
less that $J$.  These they associate with a non-Stoner continuum and 
speculate that this continuum comes down to zero energy in the 
thermodynamic limit.  In contrast, Golosov \cite{8} using the large 
$S$ expansion up to fourth order claims that magnon-electron 
scattering does {\it not\/} give rise to magnon damping this implying 
that the magnons do not enter a continuum at low energies.

Mathematically the present approach is based upon an expansion in $x$ 
the concentration of carriers.  For small $x$, $S=\frac{1}{2}$ and one 
dimension our results for the magnon dispersion agree well with those 
of exact diagonalization.  Useful results are obtained for $x$ as 
large as $3/16$ even for $S=1/2$.  The magnons {\it can\/} exchange 
energy and momentum which the charge excitations and all finite $\vec 
q$ excitations have a non-zero relaxation rate confirming the 
existence a low energy non-Stoner continuum.  For $q < k_{F}$, the 
Fermi wave vector, relaxation is reduced by phase space 
considerations.  A rapid onset of relaxation will therefore occur 
around $q \approx k_{F}$.  Although most of the results presented here 
are for one dimension, these principal conclusions are also valid for 
higher dimensions and some such results will be given.  The present 
approach does not include orbital degeneracy nor phonons and the 
authors recognize that these are important physical ingredients for 
real materials.  The method is easily generalized to include such 
effects and might form the basis for a comprehensive theory at least 
for modest $x$ values.

Consider a single spin deviation relative to the ferromagnetic ground 
state for an odd number of electrons.  In the $J \to \infty$ limit, 
for a given site $n$, there are only four states of interest.  There 
is the state $\left| S, 0 \right\rangle$ with $S_{z} = S$ and no 
electron and which is considered to be the ``vacuum''.  The state 
$\left| S - 1, 0\right\rangle$ has the spin deviated and is mapped to 
a state $d^{\dagger}_{n}|\rangle$ which has a $d$-boson created on 
this vacuum.  An un-deviated state with an up conduction electron is 
$\left| S, \uparrow \right\rangle$ and maps to the $c$-fermion state 
$c^{\dagger}_{n}|\rangle $.  Lastly, a state which has a total spin 
$S+\frac{1}{2}$ but $S_{z} = S -\frac{1}{2}$ is 
$\frac{1}{\sqrt{2S+1}}\left[ \sqrt{2S} \left| S-1.  \uparrow 
\right\rangle + \left| S, \downarrow \right\rangle \right]$ with 
$t^{\dagger}_{n}|\rangle$ as its map.  The particles are ``hard 
core'', i.e., the total number \(Q_{n} = n^{c}_{n} + n^{d}_{n} + 
n^{t}_{n} \le 1\), where, e.g., \(n^{t}_{n} = 
t^{\dagger}_{n}t^{}_{n}\).  The total number of conduction electrons 
\(\hat N = \sum_{n} \left(n_{n}^{c} + n_{n}^{t} \right)\) does not 
involve \(d^{\dagger}_{n}\).
 
Subject to \(Q_{n}\le 1\) the low energy part of the Hamiltonian maps 
{\it exactly\/} to,
\begin{eqnarray}
 &&{\cal H} = -  \sum_{nn^{\prime}}t_{nn^{\prime}} ( \tilde 
 c^\dagger_{n^{\prime}} \tilde c^{}_{n} + \frac{1}{2S+1} \tilde 
 t^{\dagger}_{n^{\prime}} \tilde t^{}_{n} + H.c.) \nonumber \\
&+&
\sqrt{\frac{2S}{2S+1} }\sum_{nn^{\prime}}
t_{nn^{\prime}} (t^\dagger_{n^{\prime}}  d^{}_{n^{\prime}} c^{}_{n}
+
c^\dagger_{n^{\prime}}  d^{\dagger}_{n} t^{}_{n}  + H.c.).
\label{deux}
\end{eqnarray}
where, e.g., $\tilde c^{}_{n} = c^{}_{n} (1-n^{t}_{n} - n^{d}_{n})$.  
When \({\cal H}\) acts on a state which satisfies \(Q_{n}\le 1\), the 
result is also consistent with this constraint, i.e., this 
representation of the Hamiltonian does not mix physical with 
unphysical states.

The Hamiltonian is written as \({\cal H}^{\prime} -\mu \hat N + I_{c}\) where 
\(I_{c}= t \sum_{n} [(\tilde c^\dagger_{n^{\prime}} \tilde c^{}_{n} - 
c^\dagger_{n+1} c^{}_{n})+ (1 / (2S+1)) (\tilde t^{\dagger}_{n+1} \tilde 
t^{}_{n} - t^{\dagger}_{n+1} t^{}_{n}) \nonumber + H.c.]\) comprises 
the operators which enforce the constraint.  Then the Fourier 
transform of,
\begin{eqnarray}
 {\cal H}^{\prime}-\mu \hat N &=& \nonumber \\
  \sum_{\vec k}(\epsilon_{\vec k} &-& \mu) 
c^\dagger_{\vec k} c^{}_{\vec k}
+
\frac{1}{2S+1} \sum_{\vec k}\left(\epsilon_{\vec k} - \mu\right) t^\dagger_{k} t^{}_{k}
\nonumber \\
-
\sqrt{\frac{2S}{2S+1} }&&\hskip -20pt \frac{1}{N}  \sum_{k,q} \epsilon_{\vec q}
(t^\dagger_{\vec k+\vec q}  d^{}_{\vec k} c^{}_{\vec q}
+
c^\dagger_{\vec q}  d^{\dagger}_{\vec k} t^{}_{\vec k+\vec q}  + 
H.c.),
\label{trois}
\end{eqnarray}
where $\epsilon_{\vec k} = -\sum_{j}t_{ij}e^{i\vec k\cdot (\vec 
r_{j} - \vec  r_{j})}$. The most important part of \(I_{c}\) is,
\begin{equation}
\frac{1}{N^2} \sum_{\vec k \vec k^{\prime}, \vec q}
\epsilon_{\vec q}
\left(
c^{\dagger}_{\vec k + \vec q} c^{}_{\vec k }
d^{\dagger}_{\vec k^{\prime} - \vec q} d^{}_{\vec k^{\prime} }
\right) + H.c.,
 \label{quatre}
 \end{equation}
since this leads to the renormalization of the conduction electron 
wave function near the deviated spin.  In the magnon dispersion 
relationship, the leading term O(\(x,t\)) is an unimportant constant.  
The O(\(x,t^{2}\)) term produces renormalization and relaxation and 
will be discussed below.

\begin{figure}[t]
\centerline{ \epsfig{file=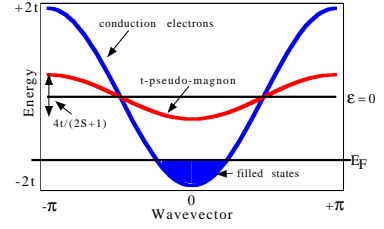,width=2.0in} } 
\vspace{10pt} 
\caption[toto]{
For one dimension, the conduction electron, or $c$-particle, band has 
width $D = 4t$ while the pseudo-magnon, i.e., $t$-fermion, band is a 
factor of $2S+1$ narrower}
\label{f1}
\end{figure}

In the zeroth order approximation, for small \(x\), the conduction 
electrons occupy the region near \(k=0\) at the bottom of a band of 
total width \(D\).  The \(t\)-particle band has a smaller width 
\([D/(2S+1)]\) and for small \(x\) this band will not have a real 
occupation, see Fig.~1.  The deviated spin, associated with the 
\(d\)-particle, has no zero order dispersion.  This becomes dispersive 
by admixture into the \(t\)-particle.  There are matrix elements, 
\(\sqrt{2S}\), of the total spin which connect the vacuum with the 
boson \(d^{\dagger}_{k}\), i.e., this is a true magnon.  There are 
also matrix elements, \(\sqrt{2S+1}\), which connect \(c^{}_{k}\) with 
\(t^{\dagger}_{k+q}\), so the finite wave-vector total spin operator,
\begin{equation}
S^{+}_{\vec q} = \sqrt{2S} d^{\dagger}_{\vec q} + 
\sqrt{2S+1}\sum_{\vec k}(t^{\dagger}_{\vec k+\vec q}c^{}_{\vec k}+ H.c.).
 \label{cinq}
 \end{equation}
For small \(x\) the dispersion in energy and momentum of the conduction
electron, \(c\)-particles, is relatively small and so, for the second
term, the  \(t\)-fermion dominates the \(q\) dispersion.  Since
it has a band width \(\sim D/S\) the \(t\)-fermion can be identified as
the {\it fast pseudo-magnon\/} with a large gap \(\sim D/2\).

\begin{figure}[t]
	\vglue -0.05in
\centerline{ \epsfig{file=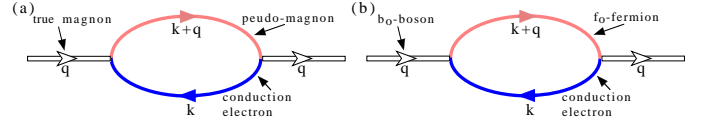,width=3.7in} } 
\vspace{10pt} 
\caption[toto]{
(a) A true, $d$-particle, magnon decays into a pseudo-magnon, i.e., a 
$t$-particle, upon absorbing a conduction electron $c$-particle. The 
momentum is conserved and each vertex is $\sqrt{2S/(2S+1)}(\epsilon_{\vec 
q}/N)$. This leads to Eqn.~(\ref{six}). (b) The equivalent for the 
ferromagnetic state. The $b_{0}$-boson, which reflects the state 
$|S,0\rangle$,  decays to a $f_{0}$-fermion, which corresponds to 
$|S,\uparrow\rangle$, upon absorbing a conduction electron. This is 
reflected by Eqn.~(\ref{huit}).}
\label{f2}
\end{figure}

The leading approximation for the dispersion of the slow magnon,
corresponds to the Feynman diagram shown in Fig.~2a, 
\begin{equation}
\omega_{\vec q}
=
\frac{2S}{2S+1}
\frac{1}{N} \sum_{\vec k} {\epsilon_{\vec k}}^{2} 
\frac{n_{\vec k}}{ \epsilon_{\vec k} - \frac{1}{2S+1} \epsilon_{\vec 
k + \vec q}}
\label{six}
\end{equation}
where \(n_{\vec k}\) is the usual conduction electron thermal 
occupation factor.  Since it involves a single sum over the occupied 
conduction electron states \(\omega_{\vec q} \sim x\) which shows that 
the \(d\)-true-magnon moves slowly for small \(x\).

It is important to confirm that the present method does not give a 
false spin gap.  In order to compare like with like, the problem with 
no reversed spins is formulated using a method in which a single site 
is treated specially so that the same approximation as used above for 
the deviated \(d\)-particle site, can be made for this special site.  
There are two, rather than a single operator for this ``impurity'' 
site, taken to be at the origin,.  The absence of a particle 
corresponds to an unphysical vacuum \(|\rangle_{0}\).  The maximum 
spin state without a conduction electron, \(|S,0\rangle\), is mapped 
to \(b^{\dagger}_{0}|\rangle_{0}\) while \(|S,\uparrow\rangle\), when 
a conduction electron is present, is reflected by 
\(f^{\dagger}_{0}|\rangle_{0}\).  The manifold with no reversed spins 
then corresponds to,
\begin{eqnarray}
&&{\cal H}-\mu \hat N =-  \sum_{nn^{\prime}}
t_{nn^{\prime}} (\tilde c^\dagger_{n} \tilde c^{}_{n^{\prime}}+ H.c.)
 -
\mu \sum_{n\ne 0}  
c^{\dagger}_{n}c^{}_{n}
\nonumber \\
&+&
 \sum_{n\ne 0} t_{0n} \left[
(f^\dagger_{0}  b^{}_{0} (c^{}_{n} + c^{}_{N-n+1})
+ H.c.  
\right] - \mu f^{\dagger}_{0}f^{}_{0} .
\label{sept}
\end{eqnarray}
With this unusual representation of the ferromagnetic ground
state it is possible to make the same approximations as with a single
deviation and a self-energy, Fig.~2b, for the \(b\)-particle, the equivalent to
the \(d\)-particle. This gives for the energy,
\begin{equation}
\omega_{0}
= \frac{1}{N} \sum_{k} {\epsilon_{\vec k}}^{2}
\frac{n_{\vec k}}{ \epsilon_{\vec k} }
= \frac{1}{N} \sum_{k} \epsilon_{\vec k} n_{\vec k},
 \label{huit}
 \end{equation}
which, surprisingly, is the exact answer.  Clearly Eqn.~(\ref{huit}) 
agrees with Eqn.~(\ref{six}) when \(\vec q=0\) which indeed confirms 
that there is no gap in the magnon spectrum.

It should also be that \(\omega_{\vec q}\) agrees with the (corrected) 
result of de Gennes for large enough \(S\).  The large \(S\) expansion 
for the {\it one dimensional near neighbor model}, i.e., with 
$\epsilon_{\vec k} \to - 2t \cos k$ gives,
\begin{equation}
\omega_{q} - \omega_{q=0}
=
\frac{1}{2S}\left[ \frac{1}{N} \sum_{k} 2t \cos k n_{k}\right] 
(1- \cos q)
 \label{neuf}
 \end{equation}
which agrees with Furukawa\cite{3}.

It is also possible to obtain an analytic result for small enough \(x\) 
independent of \(S\). Assuming that the conduction electron Fermi 
surface contracts to a point, gives,
\begin{equation}
\omega_{ q} - \omega_{q=0}
\approx
\frac{4xtS }{ 2S+1}\frac{ 1 }{ 1 - \frac{\cos q }{ 2S+1}}.
\label{dix}
 \end{equation}
For small values of \(S\), this expression gives a dispersion which 
becomes rather flat near the zone edges, see below.

\begin{figure}[t]
\centerline{ \epsfig{file=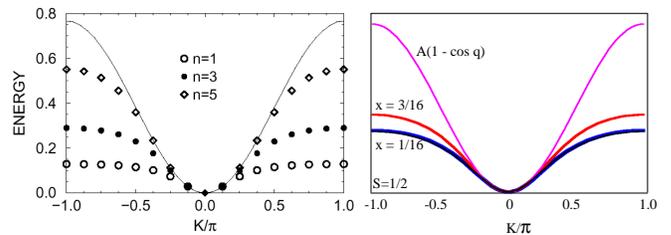,width=3.5in} } 
\vspace{10pt} 
\caption[toto]{
The right box corresponds to the present theory for concentrations \(x 
=1/16\) and \(x=3/16\) for spin \(S=1/2\).  The top curve corresponds 
to \(A(1-\cos q)\).  The scales where chosen so that all curves 
coincide at \(q =0.125 \pi\).  The small \(x\) result Eqn.~(\ref{dix}) 
cannot be distinguished from the result for \(x=1/16\).  The left box 
shows the numerical results \cite{4} with $n$ electrons and 16 sites.  
Again the scales cause the points at \(q =0.125 \pi\) to coincide with 
each other and the \(A(1-\cos q)\) solid line.  }
\label{f3}
\end{figure}

\begin{figure}[t]
	\vglue -0.08in
\centerline{ \epsfig{file=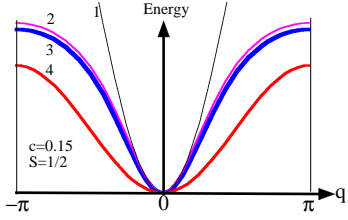,width=1.8in} } 
\vspace{10pt} 
\caption[toto]{
Curve ``3'' is Eqn.~(\ref{six}), with \(x=0.15\), for \(\omega_{q}\) 
and is almost indistinguishable from curve ``2'' which reflects the 
approximation \(\frac{4xtS }{ 2S+1}\frac{ 1 }{ 1 - \frac{\cos q }{ 
2S+1}}-2xt\).  Curve ``4'' is the large \(S\) approximation 
\(4xt[S/(S+1)^{2}](1-\cos q)\) while curve ``1'' is similar but made 
to fit the \(q^{2}\) dependence of ``3'' for small \(q\).  }
\label{f4}
\end{figure}

It is the slow \(d\)-magnon which is to be identified with the low 
energy magnon branch found in the numerical work \cite{4,5,7}.  In 
Fig.~\ref{f3} are compared the numerical results \cite{4} for 
\(S=1/2\) with the present Eqn.~(\ref{six}).  For \(x = 1/16\) and 
\(3/16\) the flattening of the dispersion near the zone edge seen in 
the numerical data is well reproduced although the difference between 
the two different concentrations is not as great as in the numerical 
data.  For \(S=1/2\) and \(x=5/16\) the Fermi surface lies in the 
\(t\)-particle band and the assumptions used for Eqn.~(\ref{six}) are 
not valid.  The result Eqn.~(\ref{six}) and the small \(x\) and the 
large \(S\) approximations are compared, for \(S=1/2\), in 
Fig.~\ref{f4}.  The small \(x\) approximation compares well with 
\(x=1/16\) even for \(S=1/2\).  In contrast the agreement with the 
large \(S\), simple \(\cos k\), result is not at all good.  The 
deviations from a simple cosine are therefore well reproduced by 
Eqn.~(\ref{dix}) and are clearly most extreme for \(S=1/2\).
  
As the small \(x\) approximation illustrates, the fattening near 
the zone edges occurs because, for small \(S\), the dispersion of the 
virtual \(t\)-particle is comparable with its energy.

Since the effective exchange increases with increasing $x$ so will the 
Curie temperature, $T_{c}$.  Larger $x$ values, Fig.~\ref{f3}, have a 
dispersion which approaches that of a Heisenberg model.  It follows 
for larger $x$ and $T_{c}$ values, the small $\vec q$ magnon stiffness 
will accord with $T_{c}$.  It is a commonplace observation that the 
zone edge softening which occurs for smaller $x$ values will cause 
$T_{c}$ to fall short of that predicted by this initial stiffness.  
This appears to be the case experimentally\cite{Dai}.

\begin{figure}[t]!
\centerline{ \epsfig{file=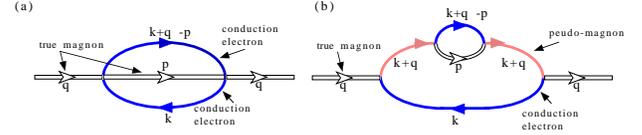,width=3.3in} } 
\vspace{10pt} 
\caption[toto]{
(a) A momentum $q$ magnon relaxes via $I_{c}$ to momentum $p$ 
transferring energy and momentum to a particle-hole conduction 
electron pair. (b) The same process but using the $t$-fermion as 
intermediate particle. }
\label{f5}
\end{figure}

It is evident from the Hamiltonian, in the form  Eqn.~(\ref{trois}), 
that there is a continuum of low energy {\it longitudinal\/}, i.e., 
charge excitations of the conduction electron \(c\)-particles.  The 
question is whether the spin excitations couple to the continuum of 
charge excitations.  This {\it is\/} the case and such a coupling 
shows up as a finite relaxation rate for the \(d\)-magnons.  In one 
dimension the leading order relaxation process for this slow magnon 
has an {\it onset\/} for relaxation which occurs only at finite 
momentum indicating that this low energy branch enters a continuum for 
finite momentum.  This relaxation corresponds to the Feynman diagrams 
shown in Fig.~\ref{f5}.  Corresponding to Fig.~\ref{f5}a is a second 
order process which involves \(I_{c}\), while Fig.~\ref{f5}b is of 
fourth order and uses the vector \(t\)-fermion in the extra 
intermediate states.  There is also interference between the two 
processes.  The net relaxation rate is of the form:
\begin{eqnarray}
&\approx& \pi \frac{{t^{\prime}}^{2}}{ N^{2}} \sum_{p^{\prime},q} n_{q}(1-n_{p+q-p})
\nonumber \\
\times
\delta(\omega_{q}&+&2t\cos k - (\omega_{p}+2t\cos (q+k- 
p)),
\label{onze}
\end{eqnarray} 
where $
t^{\prime} \approx  t + [2S / ( 2S+1)](2 t^{2} / |\mu|)$.

A particle-hole charge excitation of the conduction electron is 
created and takes away energy and momentum from the magnon.  Since 
only the initial state electron lies below the Fermi surface, such a 
process is proportional to \(x\).  The only energy scale is \(t\) 
(e.g.,\( |\mu|\sim t\)) {\it but\/} relaxation is strongly suppressed due 
to phase space considerations.  In order to understand the trends, 
consider small \(x\) and larger values of \(S\).  This implies that 
the energy dispersion of the magnon is negligible as compared to that 
of the conduction electrons.  It is not possible to conserve both 
energy and momentum for small values of the magnon \(q\) since the 
energy of the conduction electron charge excitations with the same 
momentum is too large.  Small energy but large \(q\) particle-hole 
excitations imply that the electrons scatter from \(\mp k_{F} \to \pm 
k_{F}\) and have a momentum transfer of \(\approx 2k_{F}\).  In order 
to minimize the initial magnon momentum it must scatter from \(\mp 
k_{F} \to \pm k_{F}\), i.e., the onset of magnon relaxation will occur 
for \(q \approx \pm k_{F}\) a momentum which is proportional to \(x\) 
for one dimension.  The finite magnon dispersion implied by smaller 
values of \(S\) and larger \(x\) will reduce this threshold value of 
\(q\) from this limiting value of \(\pm k_{F}\).

{\it However\/} higher order processes have a final state which still 
comprises a single magnon but now reflects a {\it pair\/} of 
conduction electron particle-hole excitations.  Even with an arbitrary 
small initial magnon momentum it is always possible to find a pair of 
particle hole excitations which, e.g., go from close to one Fermi 
surface to the other (\( \pm k_{F} \to \mp k_{F}\)) but in opposite 
senses and which have any small energy and moment.  Thus as \(x\) 
increases the pseudo-gap to momentum \(\approx \pm k_{F}\) 
progressively fills.  

Much of the above discussion generalizes to two and three dimensions. 
In particular the small $x$ approximation, 
\begin{equation}
\omega_{\vec q}
=
 \frac{2x \epsilon_{\vec k=0}S }{ 2S+1}
 \frac{1}{N} 
\frac{1 }{  1 - \frac{1}{2S+1} (\epsilon_{\vec q}/\epsilon_{\vec k=0})}
\label{douze}
\end{equation}
exhibits softening at the zone edges and relaxation will be phase 
space limited for $q < k_{F}$ even though there is no longer a 
threshold for the second order process. Evidently the orbital 
degeneracy of the real materials is of great importance. This will be 
dealt with elsewhere. 

This work was supported by a Grant-in-Aid for Scientific Research on
Priority Areas from the Ministry of Education, Science, Culture and
Technology of Japan, CREST. SEB is on sabbatical leave from the
Physics Department, University of Miami, Florida, USA and wishes to
thank the members of IMR for their kind hospitality.  SM acknowledges
support of the Humboldt Foundation.

\end{document}